\input harvmac
 \noblackbox
\baselineskip=13pt plus 2pt minus 1pt
\input epsf
\epsfverbosetrue
\def\epsfsize#1#2{\hsize}

\def\B{{\cal B}}
\def\K{{\cal K}}


%
%
\def\RF#1#2{\if*#1\ref#1{#2.}\else#1\fi}
\def\NRF#1#2{\if*#1\nref#1{#2.}\fi}
\def\refdef#1#2#3{\def#1{*}\def#2{#3}}
\def\rdef#1#2#3#4#5{\refdef#1#2{#3, `#4', #5}}

%
%
\def\ts{\hskip .16667em\relax}

\def\CMP{{\it Commun.\ts Math.\ts Phys.\ts}}

\def\FAP{{\it Funct.\ts Analy.\ts Appl.\ts}}
\def\IJMP{{\it Int.\ts J.\ts Mod.\ts Phys.\ts}}

\def\JP{{\it J.\ts Phys.\ts}}

\def\NP{{\it Nucl.\ts Phys.\ts}}
\def\PL{{\it Phys.\ts Lett.\ts}}

\def\TMP{{\it Theor.\ts Math.\ts Phys.\ts}}

\def\Zm{Zamolodchikov}
\def\AZm{A.B. \Zm}
\def\dur{H.\ts W.\ts Braden, E.\ts Corrigan, P.E. Dorey \ and R.\ts Sasaki}
%
%
\rdef\rBCDRa\BCDRa{P. Bowcock, E. Corrigan, P.E. Dorey and R. H. Rietdijk}
{Classically integrable boundary conditions for affine Toda field theories}
{\NP {\bf B445} (1995) 469}

\rdef\rBCRa\BCRa{P. Bowcock, E. Corrigan and R. H. Rietdijk}
{Background field boundary conditions for affine Toda field theories}
{DTP-95/41; hep-th/9510071}

\rdef\rBCDSc\BCDSc{\dur}
{Affine Toda field theory and exact S-matrices}
{\NP {\bf B338} (1990) 689}

\rdef\rCe\Ce{E. Corrigan}{
{\it Recent developments in affine Toda field theory},
lectures at the  CRM-CAP Summer School, Banff 1994}
{DTP-94/55; hep-th/9412213}

\rdef\rCo\Co{I.\ts V.\ts Cherednik}
{Factorizing particles on a half line and root systems}
{\TMP {\bf 61} (1984) 977}

\rdef\rCDRa\CDRa{E.\ts Corrigan, P.E. Dorey, R.H.\ts Rietdijk}
{Aspects of affine  Toda field theory on a half line}
{Suppl. Prog. Theor. Phys. {\bf 118} (1995) 143}

\rdef\rCDRSa\CDRSa{E.\ts Corrigan, P.E. Dorey, R.H.\ts Rietdijk and R.\ts
Sasaki}
{Affine Toda field theory on a half line}
{\PL {\bf B333} (1994) 83}

\refdef\rCDSa\CDSa{E. Corrigan, P.E. Dorey and R. Sasaki,
 \lq On a generalised bootstrap principle',
\NP {\bf B408} (1993) 579-99}

\refdef\rDGZa\DGZa{G.W. Delius, M.T. Grisaru and D. Zanon,
\lq Exact S-matrices for non simply-laced affine Toda theories',
\NP {\bf B382}  (1992) 365-408}

\rdef\rFKc\FKc{A.\ts Fring and R.\ts K\"oberle}
{Factorized scattering in the presence of reflecting boundaries}
{\NP {\bf B421} (1994) 159}

\rdef\rFKd\FKd{A.\ts Fring and R.\ts K\"oberle}
{Affine Toda field theory in the presence of reflecting boundaries}
{\NP {\bf B419} (1994) 647}

\rdef\rFKe\FKe{A.\ts Fring and R.\ts K\"oberle}
{Boundary bound states in affine Toda field theory}
{\IJMP {\bf A10} (1995) 739}

\rdef\rFSa\FSa{A. Fujii and R. Sasaki}
{Boundary effects in integrable field theory on a half line}
{Prog. Theor. Phys. {\bf 93} (1995) 1123}

\refdef\rGd\Gd{S.\ts Ghoshal,
`Boundary state boundary S-matrix of the sine-Gordon model',
{\it Int. J. Mod. Phys.} {\bf A9} (1994) 4801}

\rdef\rHa\Ha{T.J. Hollowood}
{Solitons in affine Toda field theories}
{\NP {\bf 384} (1992) 523}

\refdef\rCKa\CKa{H.S. Cho and J.D. Kim, \lq Boundary reflection matrix for
$d_4^{(1)}$ affine Toda field theory', DTP-95-23; hep-th/9505138}

\rdef\rFSWa\FSWa{P. Fendley, H. Saleur and N.P. Warner}
{Exact solutions of a massless scalar field with a relevant boundary
interaction}
{\NP {\bf B430} (1994) 577}

\refdef\rKc\Kc{V.\ts Ka\v c, {\it Infinite Dimensional Lie Algebras}
(Birkhauser 1983)}

\rdef\rKe\Ke{J.D. Kim}
{Boundary reflection matrix in perturbative quantum field theory}
{\PL {\bf B353} (1995) 213}

\rdef\rKf\Kf{J.D. Kim}
{Boundary reflection matrix for A-D-E affine Toda field theory}
{DTP-95-31; hep-th/9506031}

\rdef\rMv\Mv{A. MacIntyre}
{Integrable boundary conditions for classical sine-Gordon theory}
{{\it  J. Phys.} {\bf A28} (1995) 1089}

\rdef\rGZa\GZa{S.\ts Ghoshal and \AZm}
{Boundary $S$ matrix and boundary state in two-dimensional
integrable quantum
field theory}
{{\it Int. J. Mod. Phys.} {\bf A9} (1994) 3841}

\rdef\rMOPa\MOPa{A. V. Mikhailov, M. A. Olshanetsky and A. M. Perelomov}
{Two-dimensional generalised Toda lattice}
{\CMP {\bf 79} (1981) 473}

\rdef\rOTa\OTa{D. I. Olive and N. Turok}
{The symmetries of Dynkin diagrams and the reduction of Toda field equations}
 {{\it Nucl. Phys.} {\bf B215} (1983) 470}

\refdef\rOTb\OTb{D. I. Olive and N. Turok,
\lq The Toda lattice field theory hierarchies and zero
curvature conditions in Ka\v c-Moody algebras',
{\it Nucl. Phys.} {\bf B265} (1986) 469}

\refdef\rOTd\OTd{D. Olive and N. Turok,
\lq Local conserved densities and zero curvature conditions for
Toda lattice field theories',
 \NP {\bf B257} (1985) 277-301}

\refdef\rPZa\PZa{S. Penati and D. Zanon,
\lq Quantum integrability in two-dimensional sytems with boundary',
\PL {\bf B358} 63}

\refdef\rPRZa\PRZa{S. Penati, A. Refolli and D. Zanon,
\lq Quantum boundary currents for non simply-laced Toda theories',
IFUM-518-FT; hep-th/9510084}

\refdef\rPRZb\PRZb{S. Penati, A. Refolli and D. Zanon,
\lq Classical versus quantum symmetries for Toda theories with a non-trivial
boundary perturbation',
IFUM-522-FT; hep-th/9512174}

\rdef\rSSWa\SSWa{H. Saleur, S. Skorik and N.P. Warner}
{The boundary sine-Gordon theory: classical and semi-classical analysis}
{\NP {\bf B441} (1995) 421}

\rdef\rSSa\SSa{S. Skorik and H. Saleur}
{Boundary bound states and boundary bootstrap in the sine-Gordon model
with Dirichlet boundary conditions}
{USC-95-01; hep-th/9502011}

\rdef\rSk\Sk{R.\ts Sasaki}
{Reflection bootstrap equations for Toda field theory}
{in {\it Interface between Physics and Mathematics}, eds W. Nahm and J-M Shen,
(World Scientific 1994) 201}

\rdef\rSl\Sl{E.\ts K.\ts Sklyanin}
{Boundary conditions for integrable equations}
{\FAP {\bf 21} (1987) 164}

\rdef\rSm\Sm{E.\ts K.\ts Sklyanin}
{Boundary conditions for integrable quantum systems}
{\JP {\bf A21} (1988) 2375}

\rdef\rWa\Wa{G. Wilson}
{The modified Lax and and two-dimensional Toda lattice equations
associated with simple Lie algebras}
{{\it Ergod. Th. and Dynam. Sys.} {\bf 1} (1981) 361}

\rdef\rYa\Ya{A. Yegulalp}
{New boundary conformal field theories indexed by the simply-laced
Lie algebras}
{\NP {\bf B450} (1995) 641}

\rightline{DTP-96/1}
\rightline{hep-th/9601055}
\bigskip
\centerline{\bf Reflections}
\vskip 3pc
\centerline{E. Corrigan}
\bigskip
\centerline{Department of Mathematical Sciences}
\centerline{University of Durham}
\centerline{Durham DH1 3LE, England}
\medskip
\vskip 4pc
\noindent Talk given at the VII Regional Conference on
Mathematical Physics,
Bandar e Anzali, Iran, 15-22 October 1995.


\newsec{Introduction}

The purpose of this talk is to address a couple of simple-sounding
questions: what boundary conditions are compatible with

{\parindent = 55 pt
{(a)} Classical integrability?

{(b)} Quantum integrability?}

\noindent Much is now known about classical and quantum integrability
for field theories defined over the whole line in a two-dimensional
space-time with several classes of models (eg affine Toda theory,
or $\sigma$-models) having been particularly well studied.
On the other hand, the situation in which the
line is truncated to a half-line, or to an interval, is hardly explored
except for the choice of Neumann, or periodic boundary conditions,
respectively.
Recently, there has been renewed interest in this topic following
investigations in condensed matter physics in which boundaries
play a significant r\^ole. In particular, the sine-Gordon model
has been studied afresh, with a number of new results obtained,
notably by Ghoshal and Zamolodchikov
\NRF\rGZa\GZa\NRF\rGd\Gd\refs{\rGZa ,\rGd}, and others
\NRF\rSSWa{\SSWa\semi\SSa}\refs{\rSSWa}
(see also \NRF\rFSWa\FSWa\refs{\rFSWa}). The sine-Gordon
model is the simplest of the affine Toda field theories, based
on the root data of the Lie algebra $a_1$, and it is therefore natural
to explore the question of boundary conditions within the context of
other  models in the same class
\NRF\rFKc{\FKc\semi\FKd}\NRF\rSk\Sk\refs{\rFKc , \rSk}.

The affine Toda field theories (for a recent review, see \NRF\rCe\Ce
\refs{\rCe}) are scalar theories with Lagrangian
\eqn\Ltoda{{\cal L}_{0}={1\over 2}\partial_\mu\phi\cdot\partial^\mu\phi
-{m^2\over \beta^2}\sum_{i=0}^r n_i\, e^{\beta \alpha_i\cdot\phi}.}
\vfill\eject
In this Lagrangian $m$ and $\beta$ are two constants (which may be removed
from the classical field equations by a rescaling of the fields
and the space-time coordinates) while the important information
is carried by the set of vectors $\alpha_i$ and the set of
integers $n_i$.

The vectors $\alpha_i,\ i=1,2,\dots ,r$ are a set
of simple roots for a Lie algebra $g$, meaning that they
are linearly independent and any other root
may be expressed as an integer linear combination of these with either
all coefficients positive, or all coefficients negative. In particular,
the special root $\alpha_0$ is a linear combination of the
simple roots,
$$\alpha_0=-\sum_1^rn_i\alpha_i$$
where the choice of coefficients depends on $g$. For the $ade$
series of cases, $\alpha_0$ is always the \lq lowest' root. In terms of the
extended Dynkin diagrams classifying Kac-Moody algebras, $\alpha_0$
is  the Euclidean part of the extra, \lq affine' root. For the full list
of extended Dynkin diagrams, see the book by Kac
\NRF\rKc\Kc\refs{\rKc}. The diagrams full into two
classes: there are the $a,d,e$-type (including $a_{2n}^{(2)}$) which are
symmetric under the root transformation
\eqn\invert{\alpha\rightarrow 2{\alpha\over\phantom{^2} |\alpha|^2};}
and the others which come in pairs related to each other by \invert .
The two classes behave differently under quantisation
\NRF\rDGZa{\DGZa\semi\CDSa}\refs{\rDGZa} and, as will be seen
later, in the presence of boundary conditions.

If the field theory is restricted to a half-line (say,  $x\le 0$)
then there must be a boundary condition. In other words, the Lagrangian
must be modified and might take the form:
\eqn\Lboundary{{\cal L}_{\rm B}=\theta (-x)-\delta (x){\cal B}(\phi )}
on the assumption the boundary term depends only on the fields, not their
derivatives. (But, see also \NRF\rBCRa\BCRa\refs{\rBCRa}). As a consequence of
\Lboundary , the field equations are restricted to the region $x\le 0$
and supplemented by a boundary condition at $x=0$:
\eqn\equations{\eqalign{ x\le 0:\qquad \partial^2\phi_a&=-\sum_{i=0}^r
(\alpha_i)_a n_i\,
e^{\alpha_i\cdot\phi}\qquad\cr
x=0:\qquad\partial_x\phi_a&=-{\partial{\cal B}\over\partial\phi_a}\ .\cr}}
What choices of ${\cal B}$ are compatible with integrability?

\newsec{Classical integrability}

The question may be tackled via several routes. One way, probably the
simplest, is to consider the densities of the conserved charges which
integrate to yield the conserved charges for the full-line theory, and
discover how to modify them to preserve \lq enough' charges to maintain
integrability. The other is to develop a generalisation of the standard
Lax pair approach, including modifications arising from the boundary
condition, and to use it to investigate the charges and their
mutual Poisson brackets. The first approach already leads to some surprising
results  but the second is also needed in order to be certain that
conditions found by studying low spin charges are in fact all that are
necessary.

Clearly, adding a boundary condition effectively removes translational
invariance and it is no longer expected, therefore, that momentum should
be conserved. On the other hand, the total energy is given by
$$\widehat E=\int_{-\infty}^0dx{\cal E}_0 + {\cal B}(\phi )$$
and is easily seen to be conserved whatever the choice of ${\cal B}$ might be.
These two remarks already demonstrate that the best one could hope for is
that parity even charges (like energy) should be conserved whilst parity odd
charges (like momentum) will not be.

Integrable theories have infinitely many conserved charges, labelled by
their spins (for the affine Toda theories, the possible spins are the
exponents of the algebra modulo its Coxeter number), and it is expected
half of these (energy-like), at most, could be conserved on the half-line.
For the whole-line theory, it is convenient to think in terms of light-cone
coordinates and densities for spin $s$ satisfying,
$$\partial_\mp T_{\pm (s+1)}=\partial_\pm \Theta_{\pm (s-1)}.$$
However, on the half-line  the energy-like combinations are the
relevant choices and the quantities $\widehat P_s$ (the spin label will
continue
to be used even though Lorentz invariance has been lost) defined by
\eqn\Ps{\widehat P_s=\int_{-\infty}^0dx\left(T_{s+1}+T_{-s-1}-
\Theta_{s-1}-\Theta_{-s+1}\right)-\Sigma_s(\phi ),}
where the additional term must be chosen to satisfy
\eqn\Pcondition{T_{s+1}-T_{-s-1}+
\Theta_{s-1}-\Theta_{-s+1}={\partial\phi_a\over \partial t}
{\partial\Sigma_s\over \partial\phi_a}.}
Eq(\Pcondition ) is remarkably strong.

For low spin charges, such as occur in the $a_n^{(1)}$ theories
($s=2$ for $n>1$), and
the $a_1^{(1)}$ and $d_n^{(1)}$ cases ($s=3$), it is
straightforward to examine
\Pcondition\ directly, and some of  the details are available in
\NRF\rCDRSa\CDRSa
\NRF\rBCDRa\BCDRa\refs{\rCDRSa ,\rBCDRa}. The conclusion is the
following. For all of
these models, the boundary potential
must take the form
\eqn\boundary{{\cal B}=\sum_0^rA_ie^{\alpha_i\cdot\phi /2}}
where, {\bf either} every coefficient vanishes
(the Neumann condition) {\bf or},
every coefficient is non-zero with magnitude $2\sqrt{n_i}$, {\bf except} for
the case $a_1^{(1)}$, where the two coefficients are arbitrary.

It is tempting to conjecture that the form of the boundary potential provided
by \boundary\ is universal. This is indeed so, but the restrictions on
the coefficients are not quite applicable in every case. The Lax pair
approach reveals that the strong restrictions on the coefficients $A_i$
apply to every $ade$-type model but are not quite universal. The
second class of models, based on the non simply-laced algebras,
mentioned in the introduction allows a small amount
of freedom in the choice of boundary data (see table later). However, it is
only in the sine-Gordon case that the maximum freedom is permitted. Note
also, in most cases, setting the field to a specific value at the boundary
will not be compatible with integrability in the sense described.

Actually, even in the sinh-Gordon case there is a question of stability.
Recall the Bogomolny bound argument and consider the total energy for
a time independent solution to the theory restricted to a half-line
\NRF\rCDRa\CDRa\refs{\rCDRa}:
$$\eqalign{\widehat E&=\int_{-\infty}^0 dx \left({1\over 2}(\phi^\prime )^2 +
e^{\sqrt{2}\phi}+e^{-\sqrt{2}\phi}-2\right) +A_1e^{\phi_0/\sqrt{2}}
+A_0 e^{-\phi_0/\sqrt{2}}\cr
&=\int_{-\infty}^0 dx {1\over 2}\left(\phi^\prime -\sqrt{2}e^{\phi /\sqrt{2}}
+\sqrt{2}e^{-\phi /\sqrt{2}}\right)^2 +\int_{-\infty}^0 dx \sqrt{2}
\phi^\prime\left(e^{\phi /\sqrt{2}}-e^{-\phi /\sqrt{2}}\right) + \dots\cr
&\ge -4 +(A_0+2)e^{-\phi_0 /\sqrt{2}} +(A_1+2)e^{\phi_0 /\sqrt{2}}.\cr}$$
It is clear that the energy is bounded below provided $A_0\ge -2$ and
$A_1\ge -2$. Further details on the question of stability may be found in
Fujii and Sasaki
\NRF\rFSa\FSa\refs{\rFSa}.

The form of \boundary\ has been discovered by examining low spin charges
but there is always the possibility that some higher spin charge will violate
integrability unless further, more stringent, conditions are imposed. To
ensure compatibility with infinitely many charges it will be necessary
to adopt a different approach and to develop the Lax pair idea beyond
its formulation for the whole line.

First, the basic idea of a Lax pair
introduces a \lq gauge field' whose curvature vanishes if and only if
the field equations for the fields $\phi_a$ are satisfied. Explicitly,
for affine Toda theory, the Lax pair may be chosen to be
\NRF\rMOPa{\MOPa\semi\Wa\semi\OTa}\refs{\rMOPa},
\eqn\laxfull{\eqalign{&a_0=H\cdot\partial_1\phi /2+\sum_0^r
\sqrt{m_i}(\lambda E_{\alpha_i}-1/\lambda \ E_{-\alpha_i}) e^{\alpha_i\cdot\phi
/2}\cr
&a_1=H\cdot\partial_0\phi /2+\sum_0^r
\sqrt{m_i}(\lambda E_{\alpha_i}+1/\lambda \ E_{-\alpha_i}) e^{\alpha_i\cdot\phi
/2},\cr}}
where $H_a, E_{\alpha_i}$ and $E_{-\alpha_i}$ are the Cartan subalgebra
and the generators
corresponding to the simple roots, respectively, of the simple Lie algebra
providing the data for the Toda theory.
The coefficients $m_i$ are related to the $n_i$ by $m_i=n_i \alpha_i^2/8$.
The conjugation properties of the generators are chosen so that
\eqn\conj{a_1^\dagger  (x,\lambda )=a_1  (x,1/\lambda )
\qquad a_0^\dagger  (x,\lambda )
=a_0 (x,-1/\lambda ).}
Using the Lie algebra relations
$$[H, E_{\pm\alpha_i}]=\pm\, \alpha_i\, E_{\pm \alpha_i}\qquad
[E_{\alpha_i},E_{-\alpha_i}]=
2\alpha_i\cdot H/(\alpha_i^2),$$
the zero curvature condition for \laxfull\
$$f_{01}=\partial_0a_1-\partial_1a_0 +[a_0,a_1]=0$$
leads to the affine Toda field equations:
\eqn\todafull{\partial^2\phi =-\sum_0^r n_i \alpha_i e^{\alpha_i\cdot\phi}.}

For the purposes of the following discussion the boundary of the half-line
will be placed at $x=a$.

To construct a modified Lax pair including the boundary condition
derived from \Lboundary ,
it was found in \refs{\rBCDRa} to be convenient to
consider an additional  special point $x =b\ (>a)$ and two overlapping
regions $R_-:\ x \le (a+b+\epsilon )/2;\ $ and $R_+:\ x \ge (a+b-\epsilon
)/2$.
The second region will be regarded as a reflection of the first,
in the sense that if $x \in R_+$, then
\eqn\reflectphi{\phi (x )\equiv\phi (a+b-x ).}
The regions overlap in a small interval surrounding the midpoint of $[a,b]$.
In the two regions define:
\eqn\newlax{\eqalign{&R_-:\qquad \widehat a_0=a_0 -{1\over 2}\theta (x -a)
\left(\partial_1\phi +
{\partial\B\over\partial\phi}\right)\cdot H \qquad
\widehat a_1=\theta (a-x )a_1\cr
&R_+:\qquad \widehat a_0=a_0 -{1\over 2}\theta (b-x )
\left(\partial_1\phi -
{\partial\B\over\partial\phi}\right)\cdot H \qquad
\widehat a_1=\theta (x -b)a_1.\cr}}
Then, it is clear that in the region $x <a$ the Lax pair \newlax\ is
the same as the old but, at $x =a$ the derivative of
the $\theta$ function in the zero curvature condition enforces the boundary
condition
\eqn\boundary{{\partial\phi\over\partial x }=-{\partial{\cal B}\over \partial
\phi}, \qquad x =a .}
Similar statements hold for $x \ge b$
except that the
boundary condition at $x =b$ differs by a sign  in order to
accommodate the reflection condition \reflectphi .

On the other hand, for $x \in R_-$ and $x >a$, $\widehat a_1$ vanishes
and therefore the zero curvature condition merely implies $\widehat a_0$
is independent of $x $. In turn, this fact implies  $\phi$ is
independent of $x $ in this region. Similar remarks apply to the region
$x \in R_+$ and $x <b$. Hence, taking into account the reflection principle
\reflectphi , $\phi$ is independent of $x $ throughout the interval $[a,b]$,
and equal to its value at $a$ or $b$. For general boundary conditions, a glance
at
\newlax\ reveals that the gauge potential $\widehat a_0$ is different in the
two
regions $R_\pm$. However, to maintain the zero curvature condition over the
whole
line the values of $\widehat a_0$ must be related by a gauge transformation
on the overlap. Since $\widehat a_0$ is in fact independent of $x \in [a,b]$
on both patches, albeit  with a different value on each patch,
the zero curvature condition effectively requires the existence of
a gauge transformation $\K$ with the property:
\eqn\Kdef{\partial_0 \K =\K\, \widehat a_0(t ,b) -\widehat a_0(t ,a)\, \K .}
The group element $\K$ lies in the group $G$ with Lie algebra $g$, the
Lie algebra whose roots define the  affine Toda theory.

The conserved quantities on the half-line ($x\le a$) are
determined via a generating function $\widehat Q(\lambda )$ given by the
expression
\eqn\Qalt{\widehat Q(\lambda )={\rm tr}\left( U(-\infty ,a;\lambda )\
\K \ U^\dagger(-\infty, a ;1/\lambda )\right),}
where $U(x_1,x_2;\lambda )$ is defined by the path-ordered exponential:
\eqn\pathexp{U(x_1,x_2;\lambda )={\rm P}\exp \int_{x_1}^{x_2} dx\,  a_1 .}

To further understand the nature of $\K$, it is convenient to make a couple of
assumptions which turn out to be no more restrictive as far as
the boundary potential is concerned than the investigation of the low spin
charges. Suppose $\K$ is time independent, and also independent of $\phi$
in a functional sense. Then, \Kdef\ simplifies to
$$\K\, \widehat a_0(t ,b) -\widehat a_0(t ,a)\, \K =0$$
or, in terms of the explicit expression for $\widehat a_0$,
\eqn\Kdefa{{1\over 2}\left[\K (\lambda ),\,
{\partial\B\over\partial\phi}\cdot H\right]_+=-\,\left[\K (\lambda )
,\, \sum_0^r
\sqrt{m_i}(\lambda E_{\alpha_i}-1/\lambda \ E_{-\alpha_i})
e^{\alpha_i\cdot\phi /2}\right]_-,}
where the field-dependent quantities are evaluated at the boundary $x=a$.
Eq(\Kdefa ) is very strong, not only determining ${\cal B}$ but also $\K$
almost uniquely. The details of many solutions, including a catalogue of the
restrictions on ${\cal B}$ may be found in \refs{\rBCDRa}. Here, just
two examples will be given for $\K$, and a list of the parameter
restrictions for ${\cal B}$. For $\K$ the overall scale is a matter of
convenience only.

$$\eqalign{a_1^{(1)}&:\quad
\K (\lambda )=\left(\lambda^2-{1\over\lambda^2}\right){\rm I} + \pmatrix{&0&
\lambda A_1-{A_0\over\lambda }\cr
&\lambda A_0-{A_1\over\lambda} &0\cr}\cr
a_n^{(1)}&:\quad \K (\lambda )=I+2\sum_{\alpha >0}\prod_i C_i^{l_i(\alpha )}
\left[{(-\lambda)^{l(\alpha )}E_\alpha\over1+C\lambda^h}+{(-1/\lambda
)^{l(\alpha )}
E_{-\alpha}\over 1+C/\lambda^{-h}}\right],\cr}$$
where, in the latter expression, $C_i=A_i/2$, $C=\prod_iC_i$, each positive
root in
the sum may be decomposed as a sum of simple roots and $l_i(\alpha )$ denotes
the
number of times $\alpha_i$ appears in the decomposition, $l(\alpha )=
\sum_i l_i(\alpha )$.

As far as the boundary potential is concerned, the conjecture mentioned above
appears to be correct for the $ade$ series of models including the strongly
restricted boundary parameters. For all the others, the form of the boundary
potential is the same but the restrictions on the parameters are less severe.
The following diagrams represent the possibilities where $\pm$ is a discrete
choice, and $x$ or $y$ is a continuous parameter. The labels above represent
one set
of possibilities, those below represent an alternative set ($\epsilon =\pm$);
the Neumann condition is possible in all cases (but not Dirichlet). In very
few cases is there a possibility of continuously deforming away from the
Neumann
condition, maintaining integrability.

Finally, once $\K (\lambda )$ is determined, it is necessary to demonstrate its
compatibility with the classical $r$ matrix which itself determines the
Poisson brackets between the generating functions for the conserved charges
defined for the whole line theory (see for example refs
\NRF\rOTb{\OTb\semi\OTd}\refs{\rOTb}). Explicitly,
$$\{ U(\lambda ){\otimes ,} U(\mu )\}=\left[ r(\lambda /\mu ),
U(\lambda )\otimes U(\mu )\right],$$
where $r$ has the form
$$r(s)=\sum_ir_i(s)g_i\otimes g_i^\dagger ,$$
and
$U(\lambda )$ is defined in \pathexp .
Calculating the Poisson brackets between two charges of the form
given by \Qalt , will clearly require a consistency condition to be satisfied
involving $r$ and $\K$.
In earlier work
\NRF\rSl{\Sl\semi\Sm}\refs{\rSl}, the compatibility relation
appears as the main equation to be satisfied by $\K (\lambda )$ whereas
here, $\K$ has been determined independently via \Kdefa . The necessary
checking
has been carried out in ref\refs{\rBCDRa}, and $\K$ is indeed compatible
with $r$. The relationship between $r$ and $\K$ is one which would probably
repay further study: in a sense $\K$ is a fundamental object. In the quantum
case,
as will be seen in the next secion, there is a set of reflection bootstrap
equations which would imply the full-line S-matrix once the full set
of reflection factors are known.

\epsffile{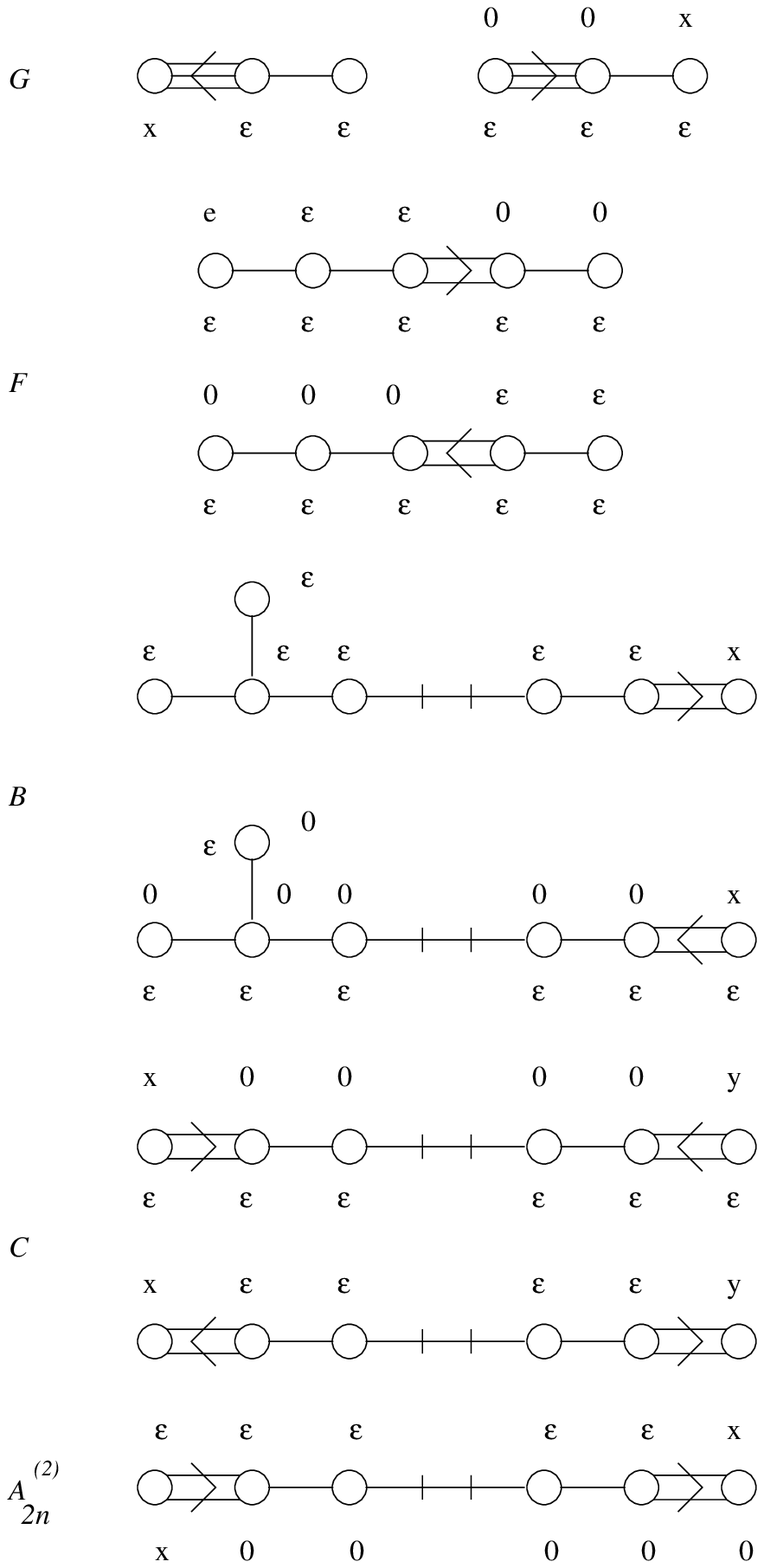}

\newsec{Quantum integrability: conjectures}

The question of quantum integrability in the presence of a boundary is more
difficult to tackle. The best one can hope to do at present is to set out
a set of hypotheses which at least allow consistent conjectures to be made
which may be checked subsequently in various ways. The main ideas were set
out by Cherednik
\NRF\rCo\Co\refs{\rCo}
many years ago and supplemented recently in work of Ghoshal and Zamolodchikov
\refs{\rGZa} concerning the sine-Gordon theory, and by Fring and K\"oberle
\NRF\rFKc{\FKc\semi\FKd}\refs{\rFKc}, and Sasaki
\NRF\rSk\Sk\refs{\rSk} concerning the affine Toda theories.

The principal idea is that particle states in the presence of the boundary at
$x=0$
continue to be eigenstates of energy and the other conserved charges. However,
an
initial state containing a single particle moving towards the boundary will
evolve into
a final state with a single particle moving away from the boundary. Thus
\eqn\quantumK{|a, v>_{\rm out}=K_{ab}(\theta )|b, -v>_{\rm in},}
where the states $a,b$ correspond to multiplets of particles distinguishable
merely via spin-zero charges, and $K_{ab}$ is a matrix which may mix the
particles
as a result of the reflection from the boundary. For real affine Toda theory
 the particles are all distinguishable and therefore there should be a set
of reflection factors, one for each particle, for each integrable boundary
condition. The velocity of  a particle is reversed on reflection, which is
equivalent to reversing its rapidity ($v=\sinh\theta$). The first major
problem is to determine a set of $K$ factors for a specified boundary
condition.
It is also supposed that particles scatter factorizably, independently of the
boundary.
This is a very strong condition which leads to the reflection Yang-Baxter
equation.
 Algebraically, the relationship is
$$K_a(\theta_a )S_{ab}(\theta_b+\theta_a )K_b(\theta_b
)S_{ab}(\theta_b-\theta_a)=
S_{ab}(\theta_b-\theta_a )K_b(\theta_b )S_{ab}(\theta_b+\theta_a)K_a(\theta_a
).$$
For affine Toda field theory where the particles are distinguishable $K$ and
$S$
are diagonal and the boundary Yang-Baxter equation is identically satisfied.

For the whole line theory there is a consistent bootstrap principle, in the
sense that
there is a consistent set of couplings between the particles, signalled by the
presence
of poles in the S-matrix at certain (imaginary) relative rapidities, and these
may be
used to relate the S-matrix elements themselves. Assuming that the whole line
couplings
remain relevant in the presence of a boundary, the bootstrap implies relations
between the various reflection factors. Algebraically, the reflection bootstrap
equation is:
\eqn\Rbootstrap{K_c(\theta_c)=K_a(\theta_a)S_{ab}(\theta_b+
\theta_a)K_b(\theta_b),}
where
$$\theta_a=\theta_c-i\bar\theta_{ac}^b\qquad
\theta_b=\theta_c+i\bar\theta_{bc}^a),$$
$\bar\theta =\pi -\theta$ and the coupling angles are the angles of the
triangle
with side-lengths equal to the masses of particles $a,b$ and $c$.

\epsffile{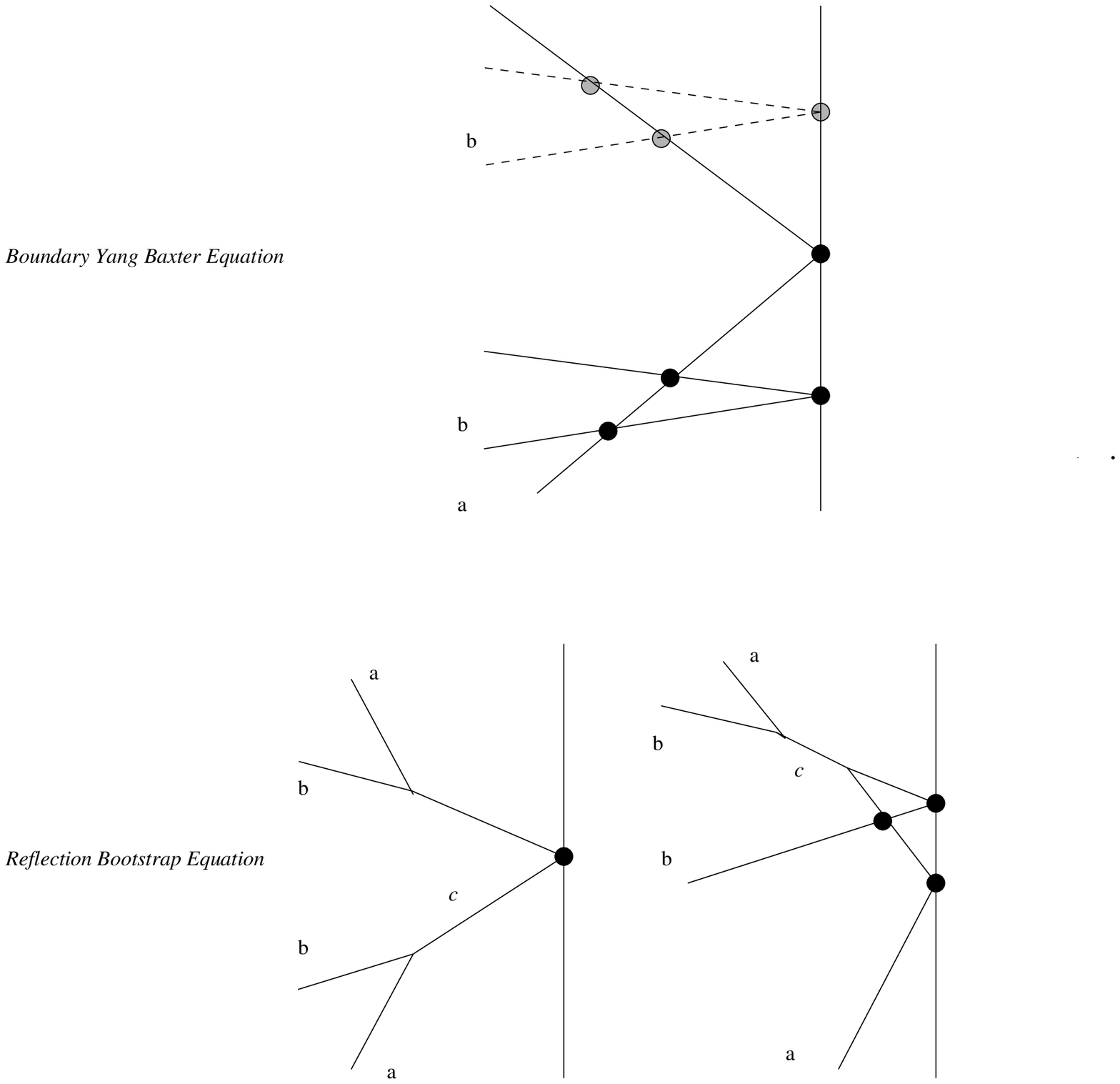}

There is also the possibility of  bound states involving a particle and the
boundary,
with their own coupling angles and bootstrap property (see
\NRF\rFKe\FKe\refs{\rGZa ,\rCDRSa , \rFKe}).

Finally, there are the Crossing relations
$$S_{ab}(i\pi -\Theta )=S_{a\bar b}(\Theta )=S_{\bar a b}(\Theta )$$
where $\Theta =\theta_b-\theta_a$, and
$$K_a(\theta-i\pi /2)K_{\bar a}(\theta+i\pi /2)S_{\bar a a}(2\theta )=1;$$
and the Unitarity relations
$$S_{ab}(\Theta )=S^{-1}_{ab}(-\Theta )\qquad K_a(\theta )=K_a^{-1}(-\theta
).$$

There are many known solutions to the reflection bootstrap equations
\refs{\rSk} but it
is not clear how to relate them to the various choices of boundary condition.
Presumably, it will be necessary to apply a semi-classical approximation,
or to use perturbation theory, although the latter may be difficult in
situations
where there are no small parameters associated with the boundary potential.
There has been some work in this direction by Kim
\NRF\rKe{\Ke\semi\CKa\semi\Kf}\refs{\rKe}, for the Neumann boundary condition.
Quantum versions of the conserved quantities have also been investigated
recently by Penati and Zanon
\NRF\rPZa\PZa
\NRF\rPRZa{\PRZa}\refs{\rPZa ,\rPRZa} whose calculations suggest
that the boundary parameters will require renormalisation. It is
notable, however, that at least in the $a_r^{(1)}$ series of cases the boundary
parameters appear to renormalise together \refs{\rPZa}.\foot{A further paper 
on the topics mentioned above, by Penati, Refolli and Zanon 
\NRF\rPRZb\PRZb\refs{\rPRZb}, has appeared since the talk was given which
suggests
that  due to the appearance of anomalies
many of the conserved charges may not survive in the quantum theory
anyhow.}

One interesting fact is the following (see \refs{\rCDRa}). If one takes the
classical
limit then it is expected that the S-matrix becomes unity. However, in the
presence of
a boundary term, the corresponding limit of the reflection factors need not be
unity.
Rather, the classical limit $K_0$ might be expected to satisfy the classical
limit
of the reflection bootstrap equation \Rbootstrap , and these classical limits
are
themselves computable via a standard linearised scattering problem involving
an \lq effective potential' determined by the (presumably static)
lowest energy solution to the classical field equations. The quantum theory
would
have to be constructed in terms of perturbations around the basic solution but
the
classical problem already has structure (including, in some cases bound states)
of a surprisingly rich kind \refs{\rCDRSa ,\rCDRa}. Unfortunately, there is
 insufficient time to discuss these matters here.

\newsec{\bf Acknowledgements}

I would like to thank the organisers of the Caspian Conference for their
kind hospitality and the opportunity to air these topics. I am also grateful
to Peter Bowcock, Patrick Dorey, Rachel Rietdijk and Ryu Sasaki for many
illuminating discussions.

\vfill\eject

\listrefs

\end